\documentclass[conference]{IEEEtran}
\IEEEoverridecommandlockouts
\usepackage{hyperref}
\usepackage{tabulary}
\usepackage{array}

\usepackage{url}
\usepackage{bm}
\usepackage{graphics}

\usepackage{subcaption}
\usepackage[font=small,skip=0pt]{caption}
\usepackage[utf8]{inputenc}
\usepackage[english]{babel}
\usepackage{pbox}
\usepackage{lmodern}
\usepackage{mathtools}
\usepackage[T1]{fontenc}

\usepackage{amssymb,amsmath,amsthm,amsfonts}
\usepackage{color}
\usepackage[ruled,norelsize,linesnumbered]{algorithm2e}
\SetAlFnt{\small}
\SetAlCapFnt{\normalsize}
\SetAlCapNameFnt{\normalsize}

\DeclarePairedDelimiter\ceil{\lceil}{\rceil}
\DeclarePairedDelimiter\floor{\lfloor}{\rfloor}

\usepackage{float}
\usepackage{cleveref}
\makeatletter
\newcommand{\removelatexerror}{\let\@latex@error\@gobble}
\makeatother

\def\BibTeX{{\rm B\kern-.05em{\sc i\kern-.025em b}\kern-.08em
		T\kern-.1667em\lower.7ex\hbox{E}\kern-.125emX}}
\begin{document}
	
\title{Improving MPI Collective I/O Performance With Intra-node Request Aggregation}
\author{\IEEEauthorblockN{Qiao Kang\IEEEauthorrefmark{1}, Sunwoo Lee\IEEEauthorrefmark{1}, Kai-yuan Hou\IEEEauthorrefmark{1}, Robert Ross\IEEEauthorrefmark{2}, Ankit Agrawal\IEEEauthorrefmark{1}, Alok Choudhary\IEEEauthorrefmark{1}, and Wei-keng Liao\IEEEauthorrefmark{1}}
	\IEEEauthorblockA{\IEEEauthorrefmark{1}
		\textit{Department of Electrical and Computer Engineering, Northwestern Univeristy} \\
		Evanston, IL, USA \\
		\{qiao.kang, slz839, khl7265, ankitag, choudhar, wkliao\}@eecs.northwestern.edu}
		\IEEEauthorblockA{\IEEEauthorrefmark{2}
		\textit{Mathematics and Computer Science Division, Argonne National Laboratory} \\
		Argonne, IL, USA \\
		rross@mcs.anl.gov}
}
\maketitle
	
\begin{abstract}
Two-phase I/O is a well-known strategy for implementing collective MPI-IO functions.
It redistributes I/O requests among the calling processes into a form that minimizes the file access costs. 
As modern parallel computers continue to grow into the exascale era, the communication cost of such request redistribution can quickly overwhelm collective I/O performance.
This effect has been observed from parallel jobs that run on multiple compute nodes with a high count of MPI processes on each node.
To reduce the communication cost, we present a new design for collective I/O by adding an extra communication layer that performs request aggregation among processes within the same compute nodes.
This approach can significantly reduce inter-node communication congestion when redistributing the I/O requests.
We evaluate the performance and compare with the original two-phase I/O on a Cray XC40 parallel computer with Intel KNL processors.
Using I/O patterns from two large-scale production applications and an I/O benchmark, we show the performance improvement of up to 29 times when running 16384 MPI processes on 256 compute nodes.
\end{abstract}
	
\begin{IEEEkeywords}
parallel I/O, MPI collective I/O, two-phase I/O, Lustre
\end{IEEEkeywords}
\section{Introduction}
The message passing interface (MPI) standard defines a set of programming interfaces for parallel shared-file access, commonly denoted as MPI-IO~\cite{MPI-3.1}.
Many large-scale scientific applications adopt MPI-IO directly or indirectly through parallel I/O libraries to obtain high I/O performance~\cite{HPF15, LiaGaoCho12, chen2009terascale, e3sm_description, SPD07}.
There are two types of MPI-IO functions: collective and independent.
The collective functions require all processes that collectively open the same shared file to participate in the calls.
Such requirement provides an opportunity for an MPI-IO implementation to coordinate processes in order to achieve better performance.
A well-known example is the two-phase I/O strategy~\cite{RosBor93A}, which has become the implementation backbone for collective I/O in almost all MPI libraries nowadays.

The two-phase I/O conceptually consists of a communication phase and an I/O phase that run one after another.
A subset of the MPI processes, defined as I/O aggregators, act as I/O proxies for the rest of the processes.
The aggregate access file region of a collective I/O call is divided among the aggregators into nonoverlapping regions, called file domains.
In the communication phase, all processes send their I/O requests to the aggregators based on the file domain assignment.
In the I/O phase, aggregators make system calls to read from or write the received requests to the file. 
The two-phase I/O strategy has been successfully demonstrated to deliver high performance on parallel machines in the past two decades.
The success of two-phase I/O strategy relies on a fast communication network by paying a relatively small cost on exchanging request data among processes to obtain in a higher gain in file system access time. 
This trade-off works effectively as the speed of I/O systems is much slower than the communication systems.
However, as the scale of parallel computers grows, soon into an exascale computing era, the number of processes running applications also increases.
The communication of the two-phase I/O exhibits an all-to-many message exchange pattern, whose cost may exceed the I/O phase for large parallel jobs, due to the high possible communication congestion on the I/O aggregators.

In this paper, we present an improvement for MPI collective I/O, referred as the two-layer aggregation method (TAM), which adds an intra-node request aggregation layer, so that the communication in the two-phase strategy consists of two layers of request aggregations.
To simply the description of the proposed method, we use a collective write as an example.
The collective read performs in the reverse order.
In the intra-node aggregation layer, MPI processes running on the same compute nodes perform a request aggregation to a subset of processes, denoted as local aggregators.
Once receiving the requests from local processes, the local aggregators coalesce them into fewer and contiguous requests.
As communication takes place among processes on the same node, the cost is expected to be relatively small.
In the meanwhile, intra-node aggregation performs on all compute nodes independently and concurrently.
Once intra-node aggregation finishes, all the local aggregators across compute nodes enter the traditional two-phase I/O to complete the collective write operation.
In contrast to local aggregators, we refer the I/O aggregators in the original two-phase I/O as the global aggregators.
An advantage of TAM is the reduction of communication congestion on the global aggregators.
In the traditional two-phase I/O, each global aggregator may receive requests from all other MPI processes, potentially causing communication contention for large-scale applications.
The intra-node aggregation alleviates such problem by breaking the all-to-many communication into two layers so that the global aggregators receive requests only from the local aggregators.
This communication contention effect is illustrated in Figure~\ref{fig:three_phase_io}, which will be discussed in details in Section~\ref{sec:design}.
In this paper, we refer the communication between local and global aggregators the inter-node aggregation.

We implement TAM in ROMIO, the implementation of the MPI-IO functions used most frequently in HPC and provided by vendors as part of their MPI implementation~\cite{thakur1997users}. 
Our performance evaluation was conducted on Theta, a Cray XC40 parallel computer with Intel KNL processors at the Argonne National Laboratory. 
Comparisons of TAM against the traditional two-phase I/O from the latest implementation of ROMIO are presented using three I/O benchmarks: E3SM-IO~\cite{e3sm_description}, S3D-IO~\cite{sankaran2006direct}, and BTIO~\cite{wong2003parallel}.
E3SM-IO and S3D-IO are I/O kernels of two large-scale production applications E3SM and S3D, respectively, while BTIO is a benchmark from NASA's NAS Parallel Benchmarks.
From the experimental results, we observed a significant time reduction in the communication costs, which contributes to end-to-end time improvement for collective I/O, ranging from 3 to 29 times faster than the traditional two-phase I/O.

\section{Related work}
Two-phase I/O is a well-known parallel I/O strategy for shared file access~\cite{del1993improved}. 
Compared with uncoordinated I/O, this strategy has successfully produced better performance for MPI collective I/O functions, as it rearranges small noncontiguous requests among processes into fewer large contiguous ones which yields shorter file access time.
The implementation of two-phase I/O in ROMIO selects a subset of MPI processes, denoted as I/O aggregators, as proxies to carry out the file access operations for the remaining processes. 
ROMIO is a widely used portable implementation of MPI-IO that adopts two-phase I/O strategy developed at Argonne National Laboratory~\cite{thakur1997users}. 
The aggregate access file region of a collective I/O call is divided among the aggregators into nonoverlapping regions, called file domains.
Each aggregator is responsible for reading and writing the file for the file domain assigned to it.
When rearranging the requests from non-aggregators to aggregators, also known as the communication phase, aggregators gather I/O requests that fall into its file domain.
The communication pattern exhibits in this phase is an all-to-many.
In the I/O phase, aggregators coalesce the received requests and write data to the file system.
For MPI collective read functions, the operation performs in a reverse order.

Many MPI libraries such as MPICH~\cite{mpich3} and OpenMPI~\cite{gabriel04:_open_mpi} adopt ROMIO as the implementation for MPI-IO functions. 
For parallel jobs running multiple MPI processes per compute nodes, ROMIO selects one aggregator per node in its default settings. 
The selection of I/O aggregators is implemented at file open time.
If the underneath file system is a Lustre, the default setting changes to select the number of I/O aggregators equal to the Lustre file striping count.
This strategy produces a ono-to-one mapping between the aggregators and the file servers (OSTs) which avoids any possible file lock conflicts and achieves the best I/O performance~\cite{liao2007implementation}. 
When the aggregate file access region is larger than the number of aggregators times the file stripe size, the two-phase I/O is carried out in multiple rounds.

Due to the fact of I/O devices being the slowest hardware component of a parallel computer system, the I/O phase is often the bottleneck of collective I/O.
Many strategies have been proposed in the past decades to reduce its cost.
A strategy to align the file domains with the file locking protocol was presented in~\cite{liao2008dynamically, liao2011design}.
Chaarawi and Gabriel proposed an algorithm that selects the number of aggregators automatically based on the file view and process topology~\cite{chaarawi2011automatically}. 
LACIO was developed as a strategy to exploit the logical I/O access pattern among processes and physical layouts of file access to optimize I/O performance~\cite{chen2011lacio}.
Pipelining the two phases in order to overlap the communication with file access was proposed in~\cite{SehSon13,TMH12}.
With the emerging of solid state driver (SSD), the cost of file access can be further reduced~\cite{zhang2013ibridge}.
Burst buffering was proposed to take advantage of faster SSD devices to improve parallel I/O performance~\cite{HAR18, BBW16}.
Although SSD cannot always replace traditional hard disks, hybrid usage of both disks by placing requests with high I/O cost to a small number of SSDs can achieve reasonably good I/O performance~\cite{he2013cost}~\cite{he2014s4d}~\cite{he2017heterogeneity}. 

MPICH-G2 presents a multi-level topology-aware strategy for MPI collective communication that can improve communication by considering reducing both intra-node and inter-node traffic \cite{KSF00}.
TAPIOCA proposes an topology-aware two-phase I/O algorithm that takes advantage of double-buffering and one-sided communication to reduce the process idle time during data aggregation \cite{TVJ17, VHM11}.
Optimizations that consider the communication topology have demonstrated their potentials for enhancing the two-phase I/O performance.

\section{Motivation}
The request rearrangement in the two-phase I/O exhibits an all-to-many communication pattern, where
for the collective write operations, all the MPI processes send their requests to the I/O aggregators.
With $P$ processes and $P_g$ I/O aggregators, there is maximum $P \cdot P_g$ send requests in each round of communication phase.
As the scale of parallel jobs increases, the value $P$ increases, but $P_g$ may stay the same.
The value of $P_g$ often depends on the file system configuration, such as file striping factor.
It is thus apparent that a much higher $P$ can cause communication congestion on the $P_g$ aggregators, causing a slow collective I/O performance~\cite{balaji2008non, hoefler2007implementation}. 


When two-phase I/O was initially proposed in the 90s, the communication phase only cost a fraction of the total execution time.
Recently, many studies have reported seeing the communication phase starts to increase to a cost that is no longer negligible, in particular on modern parallel computing systems with a massive number of computer nodes.
For example, a parallel I/O case study for E3SM climate simulation model has shown that communication phase dominates the overall performance for writing cubed sphere variables whose I/O pattern consists of a long list of small and noncontiguous requests on every MPI process~\cite{E3SM_study}. 
Given the fact that the number of MPI processes running the applications will continue to increase in the future, it is necessary to develop new communication strategies for reducing the cost of request aggregation for two-phase I/O.
Research has been conducted recently for reducing timing cost of communication phase. 
Tsujita et al. proposed a method for overlapping I/O phase with communication phase with the help of multi-threading~\cite{tsujita2014multithreaded}. 
Cha and Maeng applied a node reordering approach for reducing communication cost with non-exclusive scheduling~\cite{cha2012reducing}. 

\section{Design of Two-layer Aggregation Method}
\label{sec:design}

Our proposed new method consists of three steps for a collective I/O operation: intra-node aggregation, inter-node aggregation, and I/O phase.
We describe the method using the collective write operation.
The collective read operation performs simply in reverse order.
Focusing on communication among processes running on the same node, the intra-node aggregation gathers all requests into a subset of processes, denoted as local aggregators.
During this step, there is no communication taken place across compute nodes.
Once the requests are received, the local aggregators coalesce the requests into a potentially less number of more significant contiguous requests.
The intra-node aggregation can be implemented by shared-memory access for best performance.
During the inter-node aggregation step, local aggregators send the coalesced requests to the global aggregators. 
The I/O phase is kept the same as the original two-phase I/O where the global aggregators write the received data to the file based the assigned file domains.

\begin{figure}[t]%
	\begin{center}
		\includegraphics*[width=0.9\columnwidth]{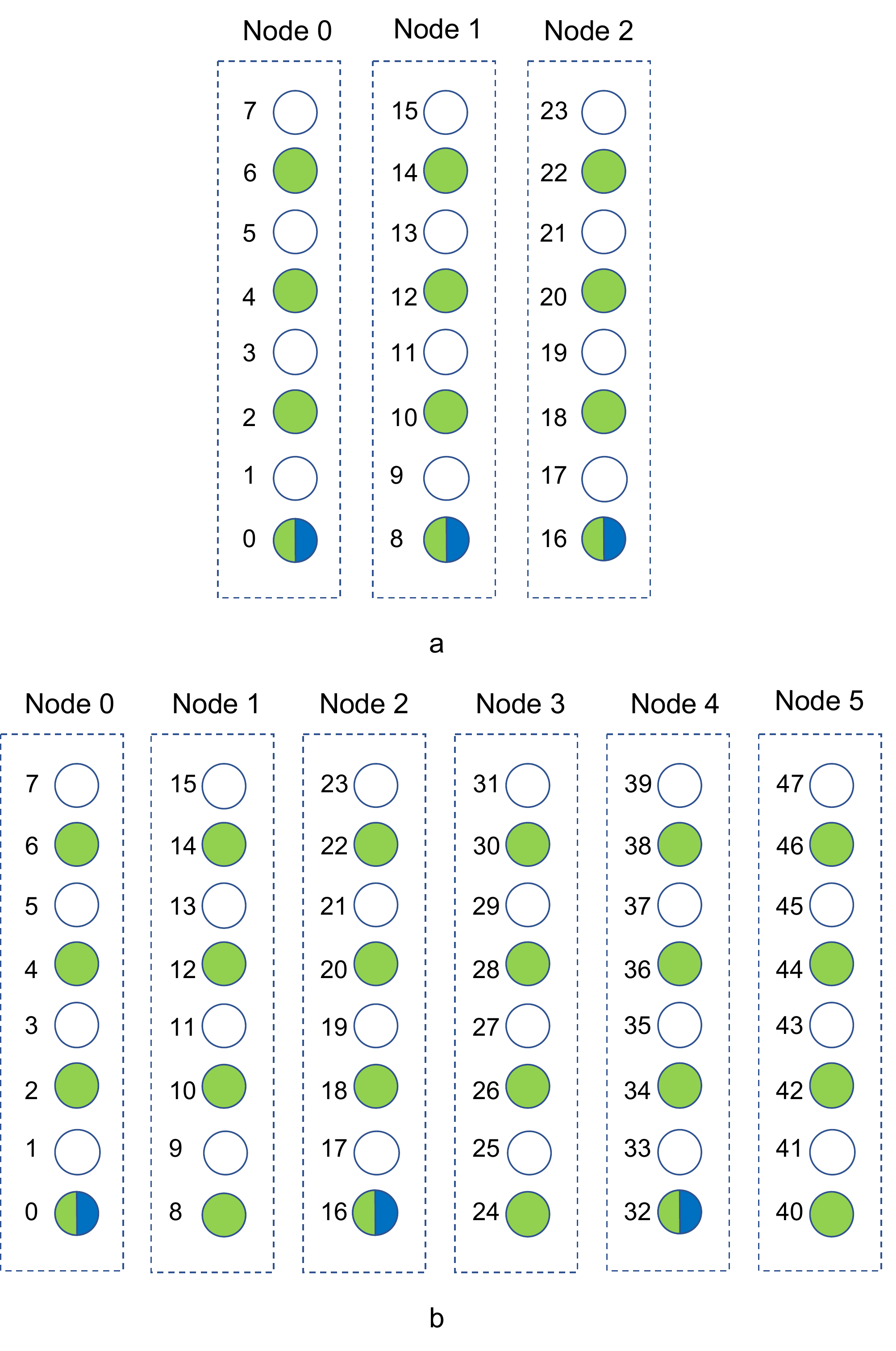}
	\end{center}
	\caption{(a) The placement of 12 local and three global aggregators on three compute nodes while 8 MPI processes are running on each node. This configuration illustrates the case when the number of nodes equals to the number of global aggregators. (b) The placement of 24 local and three global aggregators on six nodes in the case the number of nodes is more than the global aggregators. Blue circles represent global aggregators. Green circles represent local aggregators. A global aggregator can also serve as a local aggregator. The placement policy for local and global aggregators is to spread them out evenly among the available resource to prevent possible communication congestion.}
	\label{fig:aggregator_selection}
	\vspace*{-8pt}
\end{figure}

Let $P$ be the number of processes that participate in the collective I/O operation with process rank IDs ranging from $0$ to $P-1$. 
Let $L$ be a set of local I/O aggregators and $P_L$ be the number of local aggregators in total.
Let $G$ be the global aggregators selected in the original two-phase I/O and $P_G$ denotes the number of global aggregators.
We keep the number of global aggregators the same as the original two-phase I/O.

With the MPI file view mechanism, each write request from an MPI process may cover multiple noncontiguous file regions.
Individual noncontiguous file region is described by a pair of file offset and request length.
The lists of offset-length pairs must be collected by the aggregators, both local and global when calculating the message exchange amounts and numbers.
Such communication for exchanging request metadata will be required at both intra-node and inter-node aggregation steps.

\begin{figure}[t]%
	\begin{center}
		\includegraphics*[width=.47\textwidth]{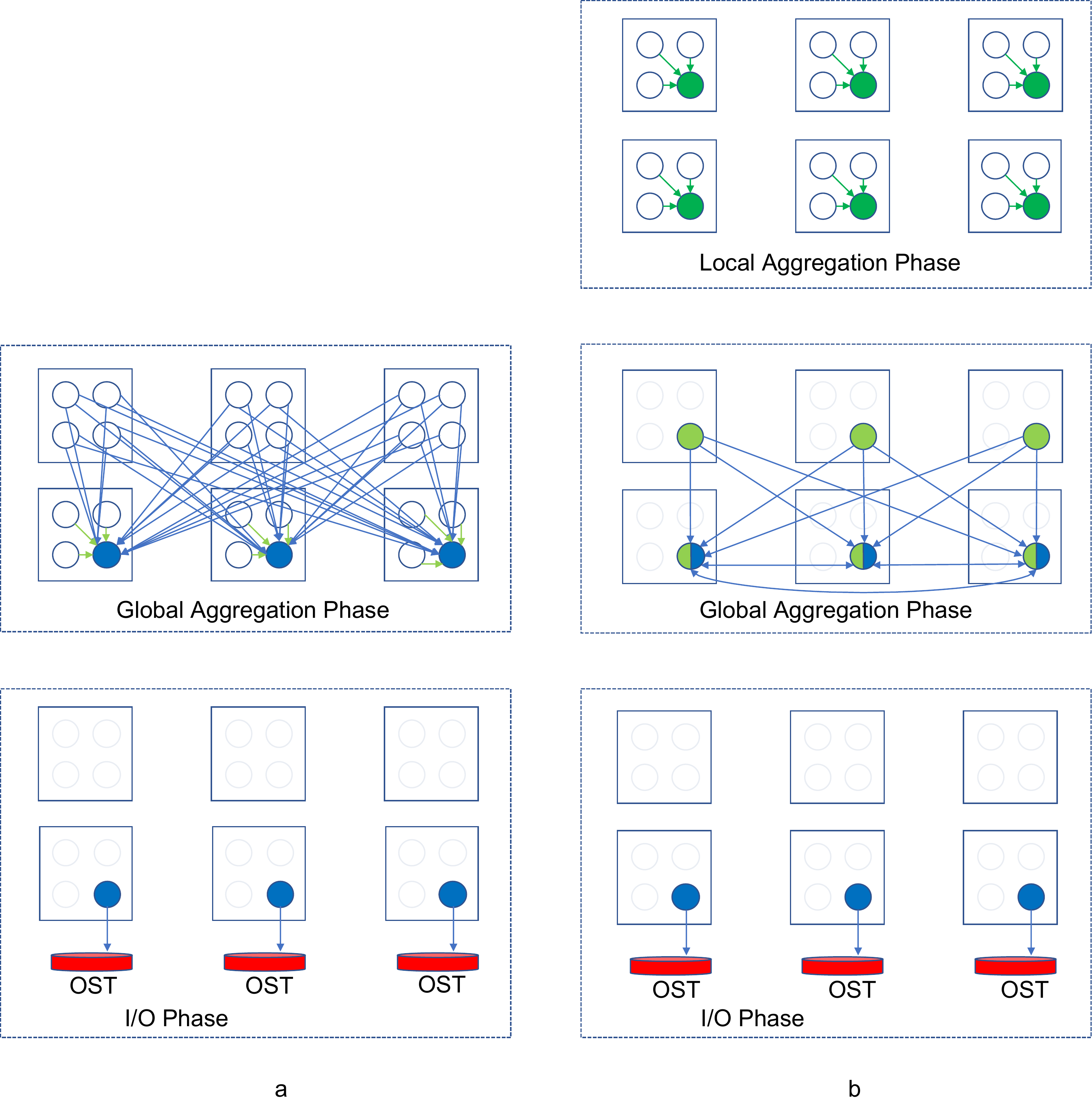}
	\end{center}  
	\caption{Illustrations of communication pattern in a collective write operation for two-phase I/O in (a) and TAM in (b). Each rectangle box represents a compute node, and circles are MPI processes. Six green circles are local aggregators, and three blues are global aggregators. Inter-node and intra-node communications are blue and green arrows, respectively. A one-to-one mapping is shown between the global aggregators and the Lustre file OST servers. In (a), the all-to-many communication pattern shows potential congestion at the global aggregators. In (b), the communication clearly shows reduced congestion at the global aggregators.}
	\label{fig:three_phase_io}
\end{figure}

\subsection{Intra-node Aggregation}

The intra-node aggregation selects a subset of local MPI processes as local aggregators.
Each local process sends its requests to one of the local aggregators.
The assignment is done by evenly dividing the local processes among the aggregators, so each aggregator receives from approximately the same number of non-aggregators.
Since each non-aggregator sends its request to only one local aggregator, the communication consists of multiple of many-to-one patterns.
There can be more than one local aggregator per compute node.
The placement of local aggregators is to spread out across all the local MPI processes.

The policy of selecting the local aggregators is described below. 
Let $q$ be the number MPI processes running on a compute node, $c$ the number of local aggregators on the node, and $e=(q\mod c)$.
We select processes with rank IDs $\ceil{\frac{q}{c}}\cdot i$ for $i=0, 1, \cdots, e-1$ and $\ceil{\frac{q}{c}}e + \floor{\frac{q}{c}}\left(i-e\right)$ for $i=e, e+1, \cdots, c$ as local aggregators. 
The selection is made at file open time. 

Figure~\ref{fig:aggregator_selection} uses two examples to illustrate the selection policy for local and global aggregators: one is when the number of compute nodes is equal to the global aggregators, and the other is when there are more compute nodes than the global aggregators.
In the former case, four out of eight MPI processes on each node are selected as local aggregators.
One of the local aggregators also serves as a global aggregator.
In the latter case, there are four local aggregators per node, but only three out of six nodes contain global aggregators, i.e., on nodes 0, 2, and 4.
Both local and global aggregators are selected by a policy evenly spreading out the processes.
This policy has been used in the two-phase I/O implementation in ROMIO, and we adopt the same for local aggregator selection.
There has been much work showing the importance of placing aggregators.
For this work, we use this simple scheme, but in principle, any existing scheme for aggregator placement for two-phase I/O implementations could be used.

A local aggregator gathers file access requests for processes with rank IDs smaller than the next local aggregator's rank and higher than or equal to its rank. 
For example, if $c = 2$ and $q = 5$. 
$r_0$ and $r_3$ are selected as local aggregators.
$r_0$ and $r_3$ gather file access requests from process group $\{r_0,r_1,r_2\}$ and $\{r_3,r_4\}$, respectively. 

In our implementation, the communication for intra-node aggregation uses asynchronous functions {\tt MPI\_Isend} and {\tt MPI\_Irecv}. 
Every process first sends the number of its flattened file access requests in offset-length pairs and the size of its write data to its local aggregator, followed by sending the write data.
Once local aggregators have gathered all file access requests, they sort file access requests into an ascending order based on the file offsets. 
Let $k$ be the average number of file access requests on every process. 
Note the offset-length pairs received from a non-aggregator are already sorted in a monotonically nondecreasing order themselves, due to the requirement by the MPI collective write \cite{MPI-3.1}.
We apply a heap merge sort algorithm to merge and sort all file access requests gathered on each local aggregator.
The time complexity of this merge sort is $O\left( \frac{P \cdot k}{P_L} \log \left( \frac{P}{P_L} \right) \right)$. 
After all file access requests are sorted, the aggregators coalesce the requests for any two consecutive requests that are contiguous. 
The proposed local aggregator selection policy may take advantage of the possibility that file access requests from processes of adjacent ranks are likely contiguous, which allow being coalesced by the same local aggregator.

\subsection{Inter-node Aggregation}
Inter-node aggregation is essentially the communication phase of the original two-phase I/O, but with participation from only the local aggregators as the write requesters.
In ROMIO, the policy for selecting global I/O aggregators when the underneath file system is a Lustre is to pick the same number of aggregators as the Lustre file stripe count.
If the number of global aggregators is higher than the number of compute nodes, there will be more than one global aggregators placed on the same compute node.
If the number of global aggregators is less than the number of compute nodes, then a subset of compute nodes is selected, and one process on each of the selected compute nodes is picked to be the aggregator.
In this case, some compute nodes do not contain global aggregators. 

Every local aggregator computes lists of its file access requests to different global aggregators based on the file domains assigned to them.
Local aggregators send their file access requests to global aggregators, which presents a many-to-many inter-node communication. 
Similar to the intra-node aggregation, global aggregators sort and coalesce the file access requests received from the local aggregators into ascending order in file offsets. 
After the end of intra-node aggregation, every local aggregator has $\frac{P \cdot k}{P_L}$ number of sorted offset-length pairs on average. 
Assuming the requests are to be evenly distributed to all global aggregators, a global aggregator will receive $\frac{P \cdot k}{P_L \cdot P_G}$ number of request from every local aggregator. 
The time complexity of merging all requests at a global aggregator is $O\left( \frac{P \cdot k }{P_G} \log\left(P_L\right) \right)$.

\subsection{I/O Phase}
In our design, the I/O phase is kept the same as the original two-phase I/O implemented in ROMIO. 
When Lustre is used, the global I/O aggregators write to the files in multiple rounds and each round an aggregator writes no more than the Lustre file stripe size.
Because the selection policy of global aggregators, a one-to-one mapping is constructed between the global aggregators and the Lustre object storage targets (OSTs), which presents possible file lock conflict and can result in a fast file access time.
The I/O phase can potentially be overlapped with the communication in the previous step, as studied in~\cite{SehSon13, TMH12}. 
However, focusing on communication improvement in this paper, we did not incorporate such enhancement technique in the design.

\subsection{Comparing with Two-Phase I/O}
The communication of TAM takes place in two stages, instead of one as in the two-phase I/O.
The communication in intra-node aggregation is $\frac{q}{c}$ to one, where $q$ is the number of MPI processes per node and $c$ is the number of local aggregators on every node.
Note $\frac{q}{c}$ is equivalent to $\frac{P}{P_L}$ if the same $q$ and $c$ are used on every node.
The inter-node communication is a many-to-many, occurring between $P_L$ local aggregators and $P_G$ global aggregators.
If implemented using asynchronous {\tt MPI\_send} and {\tt MPI\_recv} functions, the intra-node aggregation has $\frac{P}{P_L}$ number of receive operations by every local aggregator, which represents the degree of incoming message congestion. 
The inter-node aggregation has $\frac{P_L}{P_G}$ number of receives per global aggregator. 
In the traditional two-phase I/O, the communication is an all-to-many, between all $P$ processes and $P_G$ global aggregators.
Its communication phase has $\frac{P}{P_G}$ number of receives per global aggregator. 
Figures~\ref{fig:three_phase_io} (a) and (b) illustrate the communication complexity of two-phase I/O and TAM, respectively. 
It is clear to see the reduction of communication congestion on the global aggregators when TAM is used.
The two-phase I/O can be considered a special case of TAM when $P_L$ is equal to $P$.
In this case, the intra-node aggregation is skipped.

In addition to communication cost, sorting file access offset-length pairs can become a performance bottleneck if the number of the pairs is high.
For two-phase I/O, every global aggregator receives $\frac{P\cdot k}{P_G}$ number of requests on average from every process during the inter-node aggregation. 
The merge sort complexity for the traditional two-phase I/O is $O\left( \frac{P\cdot k}{P_G} \log P \right)$. 
As mentioned in the previous sections, the offset sorting for TAM is $O\left(\frac{P\cdot k}{P_G}\log\left(P_L\right)+\frac{P\cdot k}{P_L}\log\left(\frac{P}{P_L}\right)\right)$.
When $P_L\geq P_G$, TAM has a smaller time complexity in sorting.

To reduce the inter-node communication congestion on the global aggregators, the total number of local aggregators $P_L$ should be as small as possible, given the number of global aggregators $P_G$ usually being fixed to the file stripe count. 
However, if $P_L$ is too small, the communication channels available to the parallel job can be underutilized, resulting in suboptimal performance. 
On the other hand, the time spent on offset sorting at inter-node aggregation is negatively proportional to $P_L$. 
Therefore, a careful choice of $P_L$ can balance the costs of communication costs and computation on offset sorting.

\section{Experimental Results}
\begin{figure*}[t]
	\includegraphics[width=2\columnwidth]{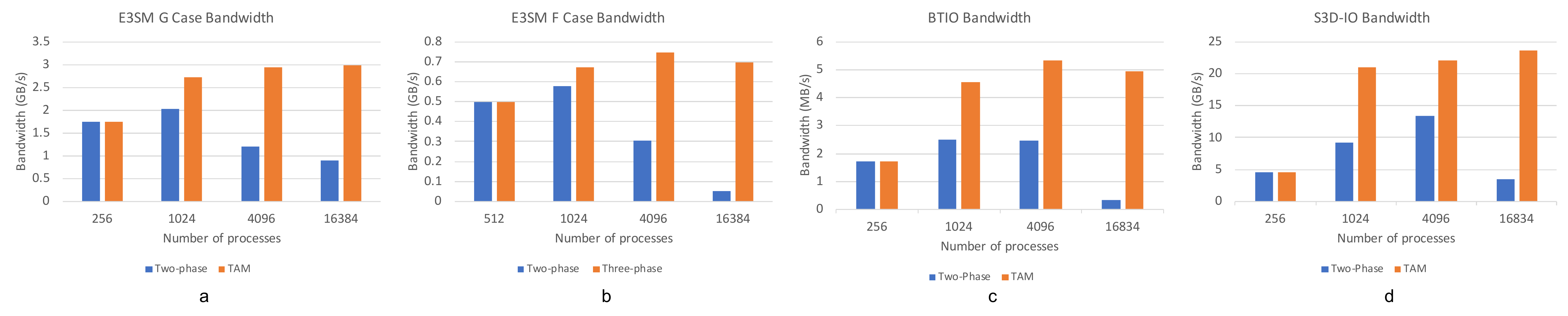}
	\caption{Write bandwidth comparisons of TAM and two-phase I/O in the strong-scaling evaluation.
		In all cases, we set the number of local aggregators to 256 for TAM.
		For the smallest runs, i.e. when the number of MPI processes is 256, TAM is equivalent to the two-phase I/O. }
	\label{fig:bandwidth_comparison}
\end{figure*}

\begin{table}[t]
	\caption{Datasets used in our evaluation.
		The second column shows the total number of noncontiguous requests.
		For E3SM F and G benchmarks, the noncontiguous requests are collected from production runs using 21600 and 9600 MPI processes, respectively.
		As we present the strong scaling results, the noncontiguous requests are partitioned among all $P$ processes used in our experiments.
		Although the numbers of requests are different among processes, the difference is small.
		Thus, the average number is shown.
		BTIO and S3D-IO have more significant numbers of noncontiguous requests and the numbers increase as the number of processes $P$.
		For S3D-IO, $y$ and $z$ are the number of processes used to partition Y and Z dimensions, respectively.}
	\label{tab:data_set}
	\centering
	\begin{tabular}{ | l | l | l | p{1 cm} |}
		\hline
		Dataset & \# noncontiguous requests & write amount \\ \hline
		E3SM G & $1.72\times10^8$ to $1.76\times10^8$ & 85GiB\\ \hline
		E3SM F & $1.35\times10^9$ to $1.37\times10^9$ & 14GiB \\ \hline
		BTIO & $512^2\times40\sqrt{P}$ & 200GiB\\ \hline
		S3D-IO & $800^2\times yz$ & 61GiB\\ \hline
		\hline
	\end{tabular}
\end{table}

We conduct our experiments on Theta, a Cray XC40 parallel computer system with Intel KNL processors at Argonne National Laboratory. 
We built a ROMIO stand-alone library out of MPICH 3.3  \cite{mpich3} and tuned it to perform at least as well as the Cray MPI compiler (Cray-MPI/7.6.0).
Running our own ROMIO library allows us to insert timers for measuring various components of the collective I/O functions.
We implemented TAM inside of ROMIO.
Using three I/O benchmarks, E3SM-IO, BTIO, and S3D-IO, we present a performance comparison between TAM and the two-phase I/O implementation in ROMIO. 
We set the Lustre the stripe size to 1 MiB and stripe count to 56, the total number of available OSTs on Theta.
All performance results are presented in a strong scaling evaluation.

Cray MPI library is not open source, so we cannot measure the two phases of a collective I/O separately.
Thus, we built ROMIO from the MPICH release version of 3.3 and adjusted the selection and placement of global I/O aggregators to ensure the end-to-end performance of a collective I/O is at least as well as the Cray MPI. 
Our first adjustment is the selection of global aggregators. 
Cray MPI selects global aggregators from different compute nodes in a round-robin fashion. 
For example, to select four aggregators from 2 nodes, each running 64 MPI processes with contiguous ranks, Cray MPI picks processes with rank IDs 0, 64, 1, and 65, in that order where 0 and 64 are located in the same node and 1 and 64 in the other same node. 
We also replaced all {\tt MPI\_Isend} with {\tt MPI\_Issend} during the aggregation.
This change is critical for large collective I/O request, where the two-phase I/O must be carried out in multiple rounds.
For non-aggregators, only asynchronous send requests were posted, and there is no receive request.
Posting asynchronous send requests using {\tt MPI\_Isend} may be cached by the operating system if the message size is small.
In this case, at the end of each round of two-phase, even though a call to {\tt MPI\_Waitall} is made, non-aggregators may continue into next round posting more asynchronous send requests, even though the send requests are still pending in the message queue.
We observed that a high number of pending asynchronous send requests could seriously hurt the communication performance, due to the possible overwhelming in the message queue processing.
Replacing {\tt MPI\_Isend} with {\tt MPI\_Issend} prevents non-aggregators from continuing into the next round of two-phase I/O, as {\tt MPI\_Issend} requires all pending send requests to be received before {\tt MPI\_Waitall} returns.
With this change, the adjusted ROMIO can sometimes outperform Cray MPI in our experiments.

\begin{figure*}[t!]
	\includegraphics[width=2\columnwidth]{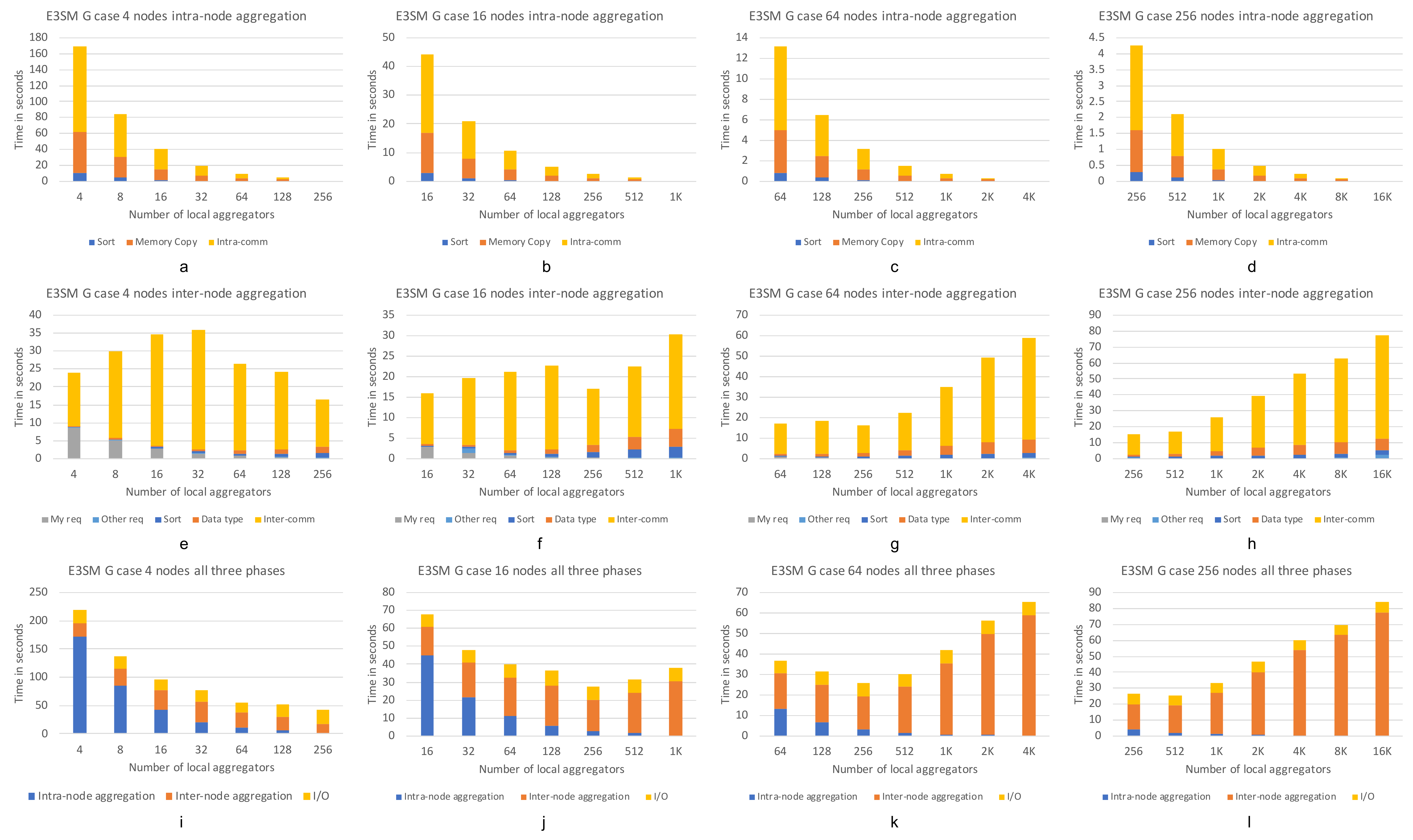}
	\caption{Timing breakdown for E3SM G case with different number of local aggregators.
		The right-most bar of all figures is the result of two-phase I/O (equivalent to TAM with $P=P_L$).
		(a) to (d) show breakdown timings for intra-node aggregation.
		(e) to (h) show breakdown timings for inter-node aggregation.
		(i) to (l) are breakdown timings of a collective write.}
	\label{fig:g_case}
\end{figure*}

In summary, TAM yields a better bandwidth performance compared with two-phase I/O and shown in Figure~\ref{fig:bandwidth_comparison}.
The total number of local aggregators $P_L$ is set to be 256 for all cases. 
This constant is empirically derived in Section~\ref{sec:e3sm_io} for the testing system and Lustre settings. 
For E3SM G and F cases, two-phase I/O drops its bandwidth as the number of processes grows. 
TAM, on the other hand, does not encounter such performance degradation. 
From Figure~\ref{fig:bandwidth_comparison} (c) and (d), we observe that TAM maintains a good write bandwidth when the number of processes increases.
The timing breakdown to be shown later presents evidence of communication cost for request aggregation does not degrade as seen in the two-phase I/O.
The rest of this section presents the analysis of timing breakdown for all three benchmarks used in this paper.

\begin{figure*}[t!]
	\includegraphics[width=2\columnwidth]{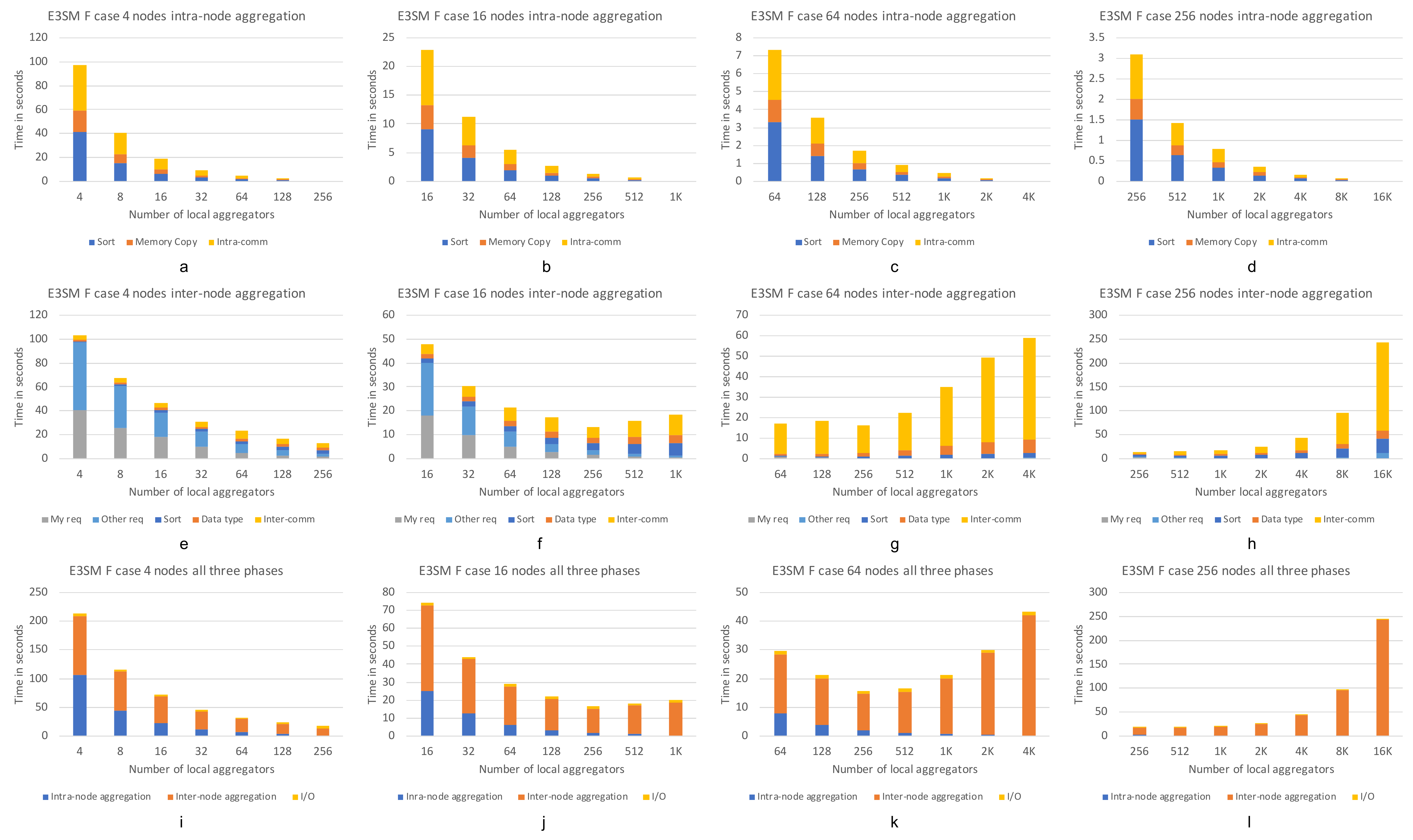}
	\caption{Timing breakdown for E3SM F case with different number of local aggregators.
		The right-most bar of all figures is the result of two-phase I/O (equivalent to TAM with $P=P_L$).
		(a) to (d) are breakdown timings for intra-node aggregation.
		(e) to (h) are breakdown timings for inter-node aggregation.
		(i) to (l) are breakdowns breakdown timings of a collective write. }
	\label{fig:f_case}
\end{figure*}

\subsection{E3SM IO Case Study}
\label{sec:e3sm_io}
E3SM~\cite{e3sm_description} is an exascale earth system modeling program for simulating atmosphere, land, and ocean behavior in high resolution. 
It is an I/O module developed to use PIO library~\cite{pio}, which is built on top of PnetCDF~\cite{pnetcdf}.
Writing data at each checkpoint is through posting nonblocking PnetCDF APIs that allow the requests to be flushed together at the end.
PnetCDF library is a high-level parallel I/O library built on top of MPI-IO.
Flushing pending nonblocking requests are implemented by aggregating the request data and combining the MPI file views before making a single call to the MPI collective write function.
In our experiments, the cost of posting the nonblocking APIs was negligible, and thus, we measured the timing of the collective write call inside the PnetCDF flush API.

We evaluated two data decompositions used in E3SM production runs,  namely F and G cases~\cite{cime}. 
F case has atmosphere, land, and runoff models components. 
The data access pattern in the F case consists of 1.36 billion noncontiguous write requests and a total write amount of 14 GiB. 
The G case has active ocean and sea ice components. 
G case decomposition file has MPAS grid data structure that consists of pentagons and hexagons on top of a spherical surface. 
The I/O pattern of G case contains a shorter list of noncontiguous requests, 180 million in total, and write amount of 85 GiB.
The data decomposition from the production runs is recorded for every process individually.
In both F and G cases, the numbers of noncontiguous requests are different among processes, but the difference is small. 
To present the performance in a strong scaling, we assign the decompositions among processes evenly when the number of processes is different from the one used in production runs.
The assignment is based on the unit of process, not the number of noncontiguous requests.
The I/O kernel of E3SM has been extracted for I/O study~\cite{E3SM_study}.

Figures~\ref{fig:f_case} and~\ref{fig:g_case} show the measured performance of TAM in breakdown timing. 
For intra-node aggregation, three major components are contributing to the timing. 
The first component is the communication for gathering request data and file access offsets to local aggregators. 
The communication pattern is many-to-one, so the total number of MPI send requests is $P$ to be received by $P_L$ local aggregators.
Each of the $\frac{P}{P_L}$ gathering operations within a node can thus run simultaneously.
The second component is to merge-sort the request file offsets at every local aggregator. 
The third component is memory operation for moving the request data into a contiguous space based on the sorted offsets. 
All three components have timings proportional to request data amount and the number of offsets. 
Therefore, as we increase the number of local aggregators $P_L$, the data amount and the number of offsets per aggregator decrease, so the time of intra-node aggregation decreases proportionally.
Figures~\ref{fig:g_case} and~\ref{fig:f_case} match such expectation from 4 to 256 nodes for both G and F cases.

For inter-node aggregation, five major components are contributing to its timing. 
The first two components are calculating local request by flattening the MPI fileview into a list of offset-length pairs and calculating others' requests to identify the global aggregators who are responsible for writing the request.
These two components are implemented in ROMIO internal functions {\tt ADIOI\_LUSTRE\_Calc\_my\_req} and {\tt ADIOI\_Calc\_others\_req} when the underneath file system is Lustre. 
In TAM, only local aggregators make calls to {\tt ADIOI\_LUSTRE\_Calc\_my\_req}. 
The time complexity is proportional to the number of locally aggregated file access requests. 
{\tt ADIOI\_Calc\_others\_req} involves a many-to-many inter-node communication from local aggregators to global aggregators. 
Its timing depends on the number of noncontiguous requests and the number of MPI requests, $P_L\cdot P_G$. 
The third component is to merge-sort the offsets of the locally aggregated file access requests.
A heap merge sort algorithm does this at every global aggregator. 
The time complexity of this component is proportional to the number of offsets. 
The fourth component is the construction of MPI derived data types at every global aggregator for receiving messages from local aggregators.
It is time complexity is also proportional to the number of local aggregators since a new MPI data type must be created to receive a message from every local aggregator into a noncontiguous memory buffer.
Using MPI derived data types to receive messages avoids additional memory operations for moving the messages into a contiguous buffer used to write to the file. 

As the number of local aggregators increases, the many-to-many communication cost in inter-node aggregation increases, because of the number of MPI {\tt isend} and {\tt irecv} requests increases.
However, in the strong-scaling study, the total write amount stays the same regardless of the number of processes. Thus, the cost of I/O phase is expected to stay the same.
In fact, in all E3SM cases shown in Figures~\ref{fig:g_case} and~\ref{fig:f_case}, the I/O phase takes only a small part of the end-to-end time.
The deterministic factor for collective write performance for E3SM cases is communication.
If the total number of MPI {\tt isend} and {\tt irecv} requests is small, we may under-utilize the available hardware communication channels, which can result in suboptimal performance. 
On the other hand, if the total number of MPI {\tt isend} and {\tt irecv} requests is too large, we may over-utilize the hardware communication channels. 
In the original two-phase I/O, every global aggregator must receive messages from all other processes.
As the number of processes increases, the communication congestion is expected to become worse at the global aggregators.
The similar problem can also happen to TAM if the number of local aggregators is enormous.
Therefore, $P_L$ should be carefully chosen to ensure excellent communication performance.

The time complexity of different components in the inter-node aggregation also depends on the number of local aggregators. 
Creating MPI derived data types also has a timing proportional to $P_L$, but in E3SM, this component costs relatively small. 
The communication components for intra-node aggregation decrease their time complexity linearly when $P_L$ increases. 
On the contrary, the communication for inter-node aggregation increases proportionally with $P_L$.
We denote the time for intra-node and inter-node aggregations as $f(P_L)$ and $g(P_L)$, respectively.
In order to decide a good choice of $P_L$, our goal is to minimize $g(P_L)+f(P_L)$. 
According to our experimental results shown in Figures~\ref{fig:f_case} (e), (f), (g), and (h) for F case and Figures~\ref{fig:g_case} (e), (f), (g), and (h) for G case, setting $P_L=256$ yields the best communication performance, among different values of $P_L$ tried in our evaluation.


\begin{figure*}[t!]
	\includegraphics[width=2\columnwidth]{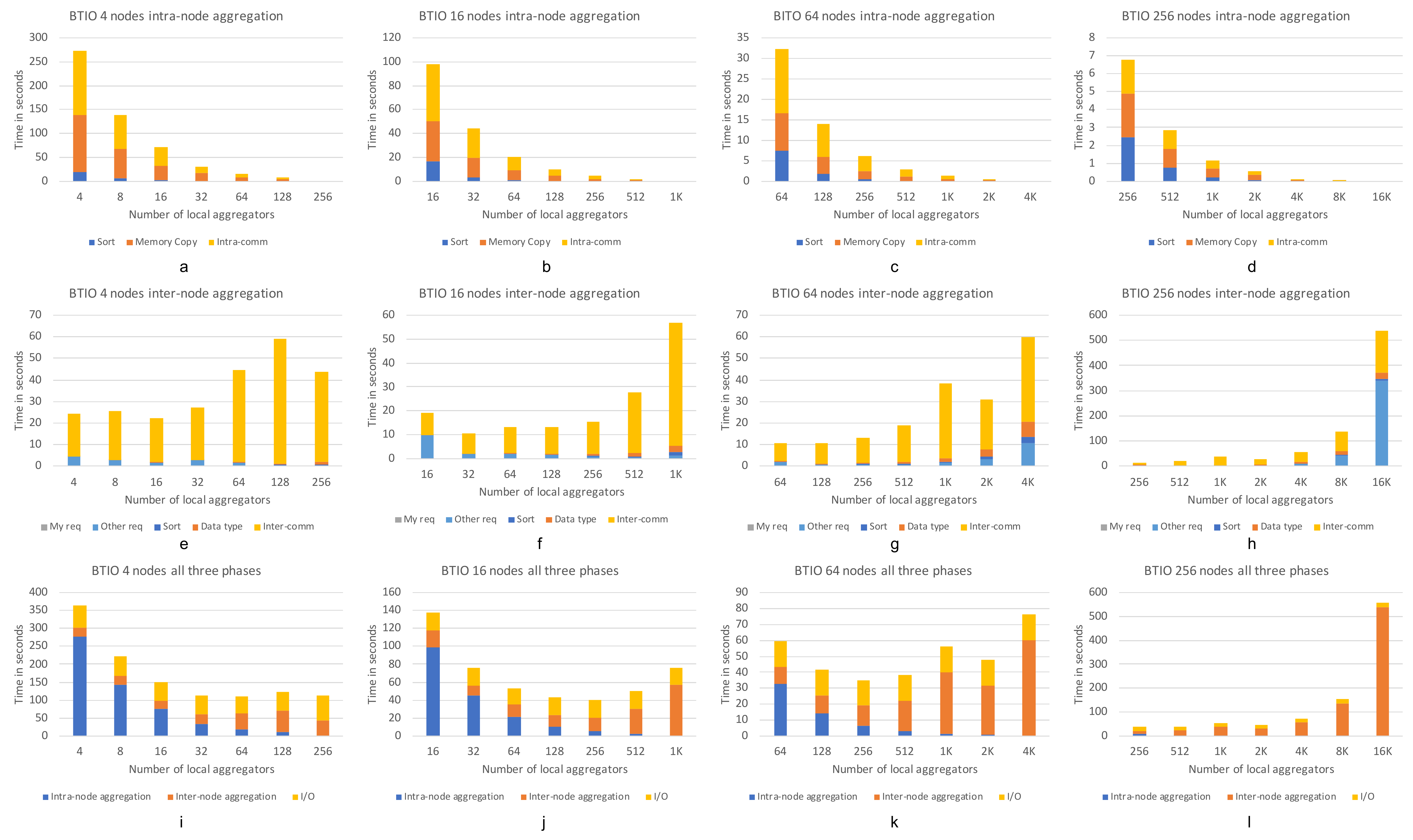}
	\caption{Timing breakdown for BTIO with different number of local aggregators.
		The right-most bar of all figures is the result of two-phase I/O (equivalent to TAM with $P=P_L$).
		(a) to (d) are breakdown timings for intra-node aggregation.
		(e) to (h) are breakdown timings for inter-node aggregation.
		(i) to (l) are breakdowns breakdown timings of a collective write. }
	\label{fig:btio}
\end{figure*}
\begin{figure*}[t!]
	\includegraphics[width=2\columnwidth]{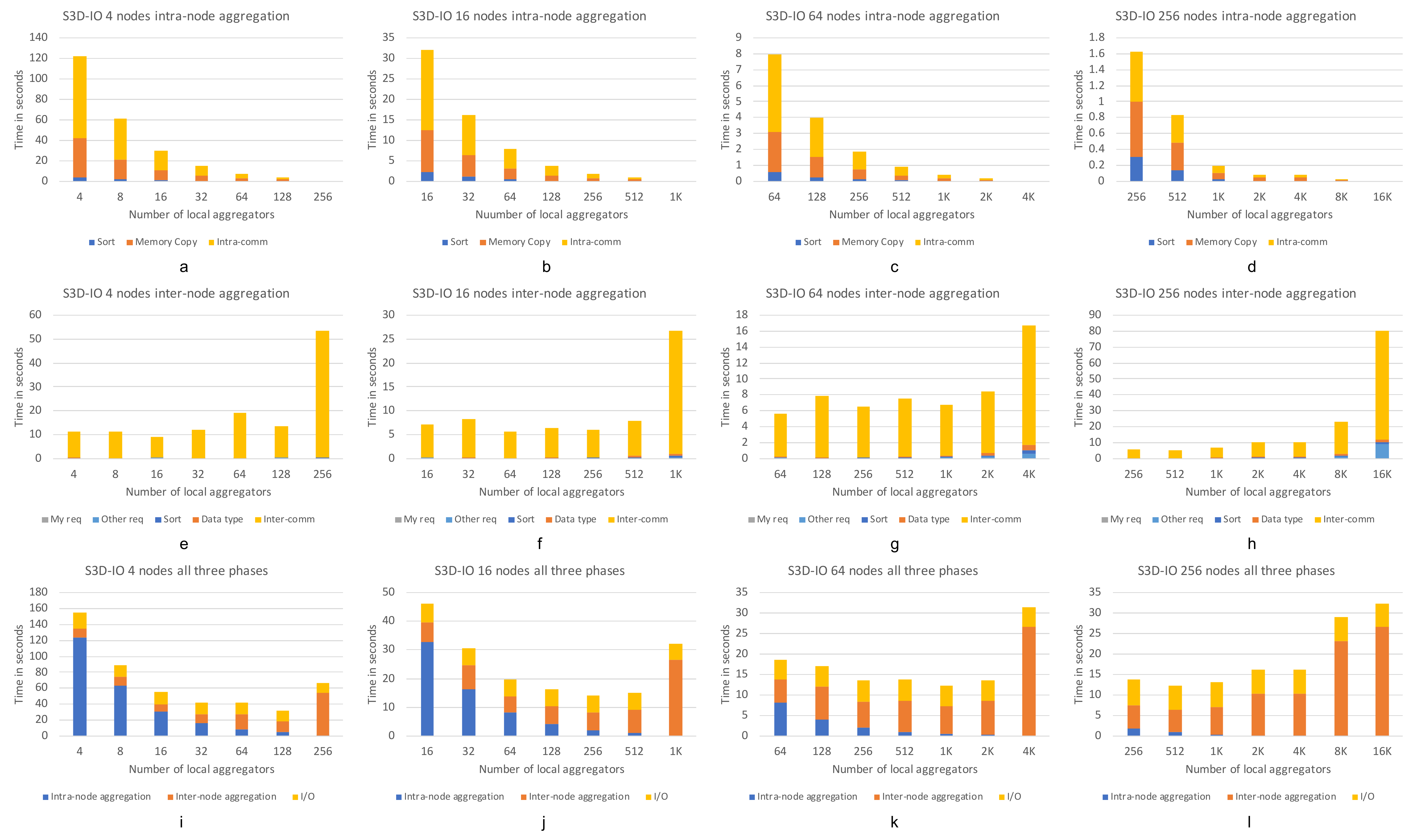}
	\caption{Timing breakdown for S3D-IO with different number of local aggregators.
		The right-most bar of all figures is the result of two-phase I/O (equivalent to TAM with $P=P_L$).
		(a) to (d) are breakdown timings for intra-node aggregation.
		(e) to (h) are breakdown timings for inter-node aggregation.
		(i) to (l) are breakdowns breakdown timings of a collective write. }
	\label{fig:s3dio}
\end{figure*}

\subsection{BTIO Benchmark}

Developed by NASA Advanced Supercomputing Division, BTIO is part of the benchmark suite NPB-MPI version 2.4 for evaluating the performance of parallel I/O~\cite{wong2003parallel}. 
BTIO uses a block-tridiagonal partitioning I/O pattern over a three-dimensional array.
The global array has five dimensions, of which the last two dimensions are not partitioned among processes. 
BTIO requires a square number of MPI processes to run, which divide the cross-sections of a global 3D array evenly. 
We set the global array size to $512\times512\times512$ array size, with 40 variables, and the fifth dimension length to 5. 
The total write amount is of size $8\times40\times512^{3}\times5\text{B}=200\text{GiB}$ with 8-byte double type per value. 
We ran 64 MPI processes per compute node and jobs allocating 4, 16, 64, and 256 nodes. 
Figure~\ref{fig:btio} shows breakdown timings of TAM. 
The right-most bars on all charts with the number of local aggregators set equal to that of all processes presents the performance of traditional two-phase I/O.

As the number processes increases, each process has smaller and more fragmented file access requests. 
For two-phase I/O, we expect to write bandwidth performance degrades as the size of noncontiguous file access requests and the network congestions increase at I/O aggregators. 
The left-most bars of Figures~\ref{fig:btio} (i), (j), (k), and (l) show the original two-phase I/O fails to scale when the number of processes increases from 1K to 16K. 
These strong scaling results indicate the bottleneck for two-phase I/O is the communication for request aggregation.

Similar to E3SM benchmarks, the intra-node aggregation in TAM has timing cost negatively proportional to the number of local aggregators $P_L$, as shown in Figures~\ref{fig:btio} (a), (b), (c), and (d). 
On the other hand, the time of inter-node aggregation increases along with $P_L$. 
In Figures~\ref{fig:btio} (e), (f), (g), and (h), calculating whether requests from all other processes fall into its file domain on each global aggregator, i.e., {\tt ADIOI\_Calc\_others\_req}, becomes significant in two-phase I/O as the number of nodes increases. 
In addition to communication congestion at global aggregators, this calculation is not negligible due to the increasing total number of non-contiguous requests being 335,544,320, 671,088,640, and 1,342,177,280 for 16, 64, and 256 nodes, respectively. 
The formula for computing the total number of noncontiguous requests is $512^2\times40\sqrt{P}$, which scales with the square root of the number of processes. 
When using 256 local aggregators, a value derived from E3SM experiments in the previous section, the best end-to-end performance for a collective write is obtained, with an approximate execution time of 40 seconds and a bandwidth of more than 5 GiB/sec. 
Furthermore, after intra-node aggregation phase, the total numbers of file access requests are coalesced to 84,377,600, 43,171,840, and 23,552,000. 
The tri-diagonal pattern has a high coalesce ratio at intra-node aggregation because the cross-section dimension is partitioned among processes with adjacent MPI rank IDs. 
Thus, the communication in {\tt ADIOI\_Calc\_others\_req} has a much less number of noncontiguous file access requests. 
As a result, TAM can effectively reduce the cost of inter-node aggregation, and improve the overall performance.
Being able to reduce the number of noncontiguous requests as well as the number of communications at global aggregators, TAM effectively scales the BTIO performance for large numbers of processes. 

\subsection{S3D-IO Case Study}

S3D-IO case study is the I/O kernel of a parallel turbulent combustion application, named S3D, developed at Sandia National Laboratories \cite{sankaran2006direct}. 
The I/O kernel is a checkpoint of three-dimensional arrays, corresponding to a 3D Cartesian mesh. 
Four variables are written at every checkpoint: mass, velocity, pressure, and temperature. 
Pressure and temperature are 3D arrays while mass and velocity are 4D arrays with the fourth dimension sizes 11 and 3. 
Processes partition the first three dimensions of every variable in a block-block-block fashion.
We set the global 3D array size to $800\times800\times800$. 
This setting results in a total write amount of size $8\times\left(11+3+1+1\right)\times800^3$B = 61 GiB with 8-byte double type per value.

The block-block-block partitioning of the global 3D array is expected to produce noncontiguous requests that mostly can be coalesced at the local aggregators, a similar effect observed in the BTIO benchmark. 
The number of file access requests after merge sort is at most $\left(\frac{1}{2}\right)^{\frac{p}{P_L}}$ of the original one.
However, S3D-IO has a smaller number of noncontiguous file access requests, compared with BTIO, which is about 327,680,000 on 16,384 processes. 
The number of noncontiguous requests is $800^2yz$.
$y$ and $z$ are the numbers of processes partitioning dimensions $Y$ and $Z$, respectively. 

Figure~\ref{fig:s3dio} illustrates the breakdown timings for S3D-IO. 
The timings for intra-node aggregation match the performance behaviors observed in the BTIO and E3SM benchmarks where the timings are negatively proportional to the number of local aggregators. 
For S3D-IO, the dominant term is the inter-node aggregation. 
With a small number of local aggregators, majority of the noncontiguous file access requests can coalesce. 
Thus, we expect a small timing cost for communicating and offset sorting, which can be seen in Figures~\ref{fig:s3dio} (a), (b), (c), and (d). 
Setting the total number of local aggregators to 256, we achieved better end-to-end execution time than the two-phase I/O.

\section{Conclusion}


In this paper, we demonstrated the communication cost of two-phase I/O strategy could become the performance bottleneck for MPI collective I/O.
Adding an intra-node aggregation can effectively reduce the communication congestion at the global I/O aggregators and thus allows the collective I/O to scale to larger numbers of MPI processes.
Our experiments show that TAM works well for applications that make a large number of non-contiguous file access requests from every process.
As the HPC community is entering the exascale era, keeping MPI-IO implementation scalable to higher numbers of MPI processes has become increasingly important.
Our proposed TAM is designed to tackle the problem from the angle of reducing communication congestion on global aggregators.
There are other opportunities to improve the collective I/O performance further.
One example is to use shared memory access to conduct the intra-node aggregation, instead of calling MPI communication functions.
This approach can also reduce the memory footprint.
Another possibility is to overlap the communication with the I/O as the pipelining approaches proposed in~\cite{SehSon13, TMH12}. 
The intra-node aggregation in TAM is motivated by the scenario when the number of MPI processes per compute node is large, resulting in a high number of messages sending to a small number of I/O aggregators.
However, if the number of MPI processes per node is small, such as a common practice used in the MPI-OpenMP programming model, TAM will not be effective.
Future work includes the extension of two-level aggregation idea to consider MPI processes allocated at compute nodes that are physically near with each other sharing the same communication hardware, such as routers in the same cabins.
All in all, communication efficiency in MPI collective I/O no doubt presents a challenge soon for software developers that require to be addressed in large-scale HPC problems.

\clearpage
\bibliographystyle{ieeetr}
\bibliography{sigproc} 
\end{document}